\documentclass[referee]{sn-jnl}

\usepackage{amsmath}
\usepackage{graphicx}
\usepackage{subcaption}
\usepackage{enumitem}
\usepackage[capitalize]{cleveref}
\usepackage{todonotes}
\usepackage{booktabs}
\usepackage{tabularx}
\usepackage{soul}
\usepackage{ulem}
\usepackage{xspace}
\usepackage{apalike}
\usepackage{setspace}
\usepackage{gensymb}

\newcolumntype{C}{>{\centering\arraybackslash}X}

\newcommand{\etal}{\textit{et al.}\xspace}
\newcommand{\bd}{\textit{Bd}\xspace}

\newcommand{\frontmatter}[2]{}

\newcommand{\deidentify}[2]{#2}

\begin{document}

\title{Sunlight-heated refugia protect frogs from chytridiomycosis: a mathematical modelling study}

\author[1]{\fnm{Claire M.} \sur{Miller}} 
\affil[1]{\orgdiv{Auckland Bioengineering Institute}, \orgname{University of Auckland}, \orgaddress{\city{Auckland}, \country{New Zealand}}}
    
\author[2]{\fnm{Michael J.} \sur{Lydeamore}}
\affil[2]{\orgdiv{Department of Econometrics and Business Statistics}, \orgname{Monash University}, \orgaddress{\city{Clayton}, \state{Victoria}, \country{Australia}}}
    
\author[3]{\fnm{Jennifer A.} \sur{Flegg}}
\affil[3]{\orgdiv{School of Mathematics and Statistics}, \orgname{The University of Melbourne}, \orgaddress{\city{Parkville}, \state{Victoria}, \country{Australia}}}
    
\author[4]{\fnm{Lee} \sur{Berger}}
\author[4]{\fnm{Lee F.} \sur{Skerratt}}
\affil[4]{\orgdiv{Melbourne Veterinary School}, \orgname{The University of Melbourne}, \orgaddress{\city{Werribee}, \state{Victoria}, \country{Australia}}}

\author[5,6]{\fnm{Anthony W.} \sur{Waddle}}
    \equalcont{Joint senior author}
\affil[5]{\orgdiv{School of Natural Sciences}, \orgname{Macquarie University}, \orgaddress{\city{Sydney}, \state{New South Wales}, \country{Australia}}}
\affil[6]{\orgdiv{Applied BioSciences}, \orgname{Macquarie University}, \orgaddress{\city{Sydney}, \state{New South Wales}, \country{Australia}}}

\author*[7]{\fnm{Patricia Therese} \sur{Campbell}}
    \email{patricia.campbell@unimelb.edu.au}
    \equalcont{Joint senior author}
\affil[7]{\orgdiv{Department of Infectious Diseases}, \orgname{The University of Melbourne, at the Peter Doherty Institute for Infection and Immunity}, \orgaddress{\city{Parkville}, \state{Victoria}, \country{Australia}}}

\abstract{
\doublespacing
\begin{enumerate}[leftmargin=*]
\item The fungal disease Chytridiomycosis poses a threat to frog populations worldwide. It has driven over 90 amphibian species to extinction and severely affected hundreds more. Difficulties in disease management have shown a need for novel conservation approaches. In order to most effectively assess and deploy a new management strategy in the natural environment, it is critical to have an estimate of the effect size of the proposed intervention. 
\item We present a novel mathematical model for chytridiomycosis transmission in frogs that includes the natural history of infection, to test the hypothesis that sunlight-heated refugia reduce transmission. The model is fit using approximate Bayesian computation to experimental data where a cohort of frogs, a fixed subset of which had cleared a prior infection, were provided access to either sunlight-heated or shaded refugia. Using our model, we can estimate the extent to which prior chytridiomycosis infection protects against subsequent infection, and quantify the effect of sunlight-heating of refugia. 
\item Results estimate a 40\% reduction in chytridiomycosis transmission when frogs have access to sunlight-heated refugia, compared to shaded refugia. This strongly supports the hypothesis that the sunlight-heated refugia reduce disease transmission. 
\item Frogs that were infected and recovered were estimated to have a reduction in susceptibility of approximately 97\% compared to frogs with no prior infection. 
\item \textit{Policy implications}: This research provides quantitative evidence supporting sunlight-heated refugia as an effective disease management tool for chytridiomycosis in frog populations. By estimating both the impact of refugia and the protective effects of prior infection, the model provides an evidence base for implementing sunlight-heated refugia as part of amphibian conservation strategies. This work represents an important first step in using mathematical modeling to inform policy on the design and implementation of habitat-based interventions to support amphibian population recovery and long-term sustainability.
\end{enumerate}
}


\maketitle

\frontmatter{CRediT Statement of authorship} {CM: software, methodology, formal analysis, writing - original draft, writing - review and editing, visualisation. ML: software, methodology, formal analysis, writing - original draft, writing - review and editing. JF: conceptualization, methodology, writing - original draft, writing - review and editing. LB: conceptualization, writing - review and editing. LS: conceptualization, writing - review and editing. AW: conceptualization, investigation, writing - review and editing. PC: conceptualization, methodology, writing - original draft, writing - review and editing.}

\frontmatter{Data availability statement}{The data and code that support the findings of this study are openly available from Zenodo, at \url{http://doi.org/10.5281/zenodo.15179344} (v1.0).}

\frontmatter{Conflict of interest statment}{The authors declare no conflicts of interest.}


 
\section{Introduction}
Worldwide, amphibian populations are facing unprecedented decline due to the devastating spread of chytridiomycosis, a fungal disease primarily caused by \textit{Batrachochytrium dendrobatidis} (\bd) \cite{berger1998,scheele2019}.  Reintroduction programs are the leading approach to amphibian population recovery, but the success of such programs is hampered by the transmission of chytridiomycosis between frogs. This difficulty highlights a need to better understand the disease dynamics between frogs, and the conditions under which its transmission can be reduced.

The \bd fungus thrives under cooler conditions with temperatures exceeding 30\degree C in the laboratory typically being lethal to the pathogen
\cite{piotrowski2004,voyles2017}. In line with this observation, warmer temperatures may increase survival from \bd infection in frogs \cite{berger2004}. 
As such, the impacts of the pathogen are worse under cooler conditions but appear limited in warm habitats and seasons that provide refuge from disease \cite{berger2004,scheele2019,puschendorf11}.
In some species, recovery from infection after exposure to heat has been found to result in reduced susceptibility to future infections even under temperatures that should favor disease \cite{waddle2021,waddle24}. Therefore, wild frogs that can warm themselves and clear their \bd infections using thermal refugia may build resilience to future infections. 
In a novel field experiment, the impacts of sunlight-heated artificial refugia and heat-induced pathogen clearance (hereafter `vaccination') were shown to drastically reduce \bd infection intensity in outdoor experiments conducted over winter months \cite{waddle24}. The mechanisms behind this improvement, and the interaction between the effects of temperature and vaccination remain uncertain; however, this can be quantified using mathematical modelling. 

Mathematical modelling of chytridiomycosis infection in amphibians is limited. Early work from Briggs \etal used a dynamic compartment model with demography to show that survival of some fraction of infected frogs is needed to allow long-term persistence of the population with chytridiomycosis \cite{briggs05}. Later work by Briggs \etal used a multistate model to investigate conditions for the persistence of infected populations, finding the different population outcomes of infection may be due to density-dependent transmission between the host and the pathogen \cite{briggs2010enzootic}. Drawert \etal used a dynamic transmission model to investigate the effect of interventions including culling and anti-fungal treatment \cite{drawert2017using}. More recently, statistical modelling has highlighted the importance of including temporal infection load changes rather than infection status alone in disease models \cite{hollanders2023recovered}. There are, however, still many unanswered questions related to chytridiomycosis infection in amphibians that mathematical modelling can be used to address. Perhaps most importantly it can be used to estimate effect sizes and therefore, provide wildlife management with the information it needs to help amphibians populations to recover.

In this paper, our aim was to estimate the extent to which recovery from prior chytridiomycosis infection reduced new infections, and quantify the difference in infection rates  between frogs with and without access to sunlight-heated artificial refugia. The estimation of effect sizes is vital for understanding the potential impact of these interventions when planning threatened species conservation programs. To do this, we built a dynamic mathematical model informed by data from a laboratory experiment to fit the field experiment observations from Waddle \etal \cite{waddle24}.  In Section \ref{sec:methods}, we describe the experimental data and model construction. Then, in Section \ref{sec:results} we present the results of the statistical estimation. Finally, Section \ref{sec:discussion} presents the strengths and weaknesses of this study, and outlines plans for future work.
    
\section{Methods} \label{sec:methods}

The goal of this study was to determine the extent to which prior chytridiomycosis infection (`vaccination') reduces frog susceptibility to infections, and if infection rates vary with environment due to the temperature dependence of infection load. Environments either include access to sunlight-heated artificial refugia (`unshaded') or shade-cloth covered refugia (`shaded'). 
We hypothesise these effects would be evident through a vaccination and environment effect on the transmission rate.
Model parameters are estimated using experimental data, on the endangered green and golden bell frog (\textit{Litoria aurea}), from Waddle \textit{et al.} \cite{waddle24}.

\subsection{Mathematical Model}
\subsubsection{Transmission model}
\begin{figure}
    \centering
    \includegraphics[width=0.6\linewidth]{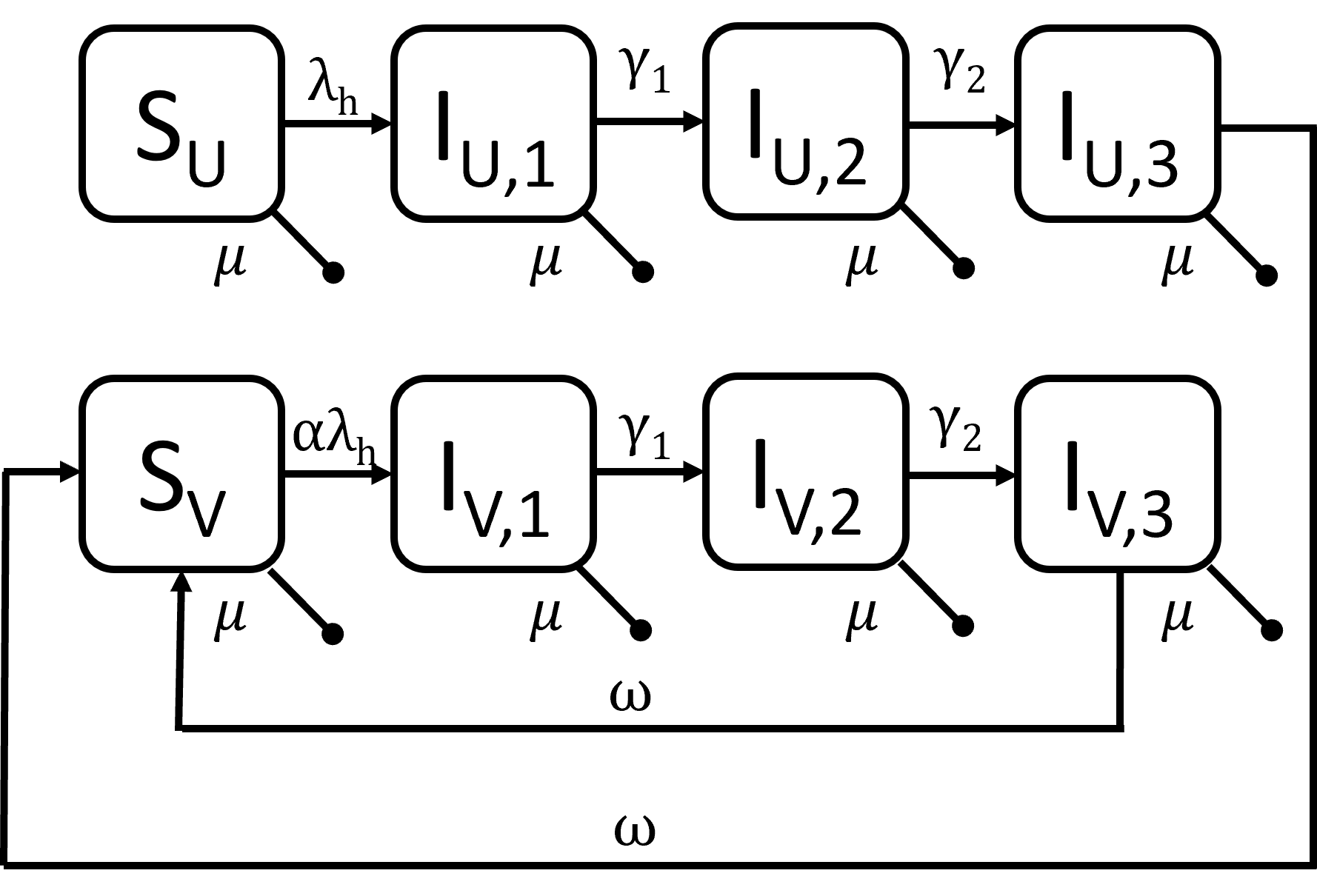}
    \caption{Schematic of epidemiological model for chytridiomycosis infection dynamics incorporating the effects of prior pathogen infection (unvaccinated and vaccinated frogs shown with subscript $U$ and subscript $V$, respectively). Unvaccinated frogs can either be susceptible ($S_U$), or in one of three states of infection ($I_{U,1},\ I_{U,2},\ I_{U,3}$). After clearance of infection, frogs are moved into $S_V$ (susceptible but recovered from prior infection) and when infected move through $I_{V,1},\ I_{V,2},\ I_{V,3}$ (infected but have prior infection). The loss rate of frogs, $\mu$, occurs from each model compartment.}
    \label{fig:model-diagram}
\end{figure}

To test our hypothesis, we built a continuous-time Markov chain (CTMC) model of chytridiomycosis infection in frogs which included prior infection status. The transmission model is summarized in \cref{fig:model-diagram}, with the full set of model transitions given in Table \ref{tab:transition-rates}.

Frogs can be either vaccinated (recovered from prior infection), denoted by $V$, or unvaccinated (no prior infections), denoted by $U$. 
Susceptible frogs become infected according to their vaccination status and the force of infection imposed on them by infectious frogs. 
After recovery from an infection, unvaccinated frogs transition into the vaccinated trajectory.
Frogs are lost from all model states due to death or not being observed, with the time-varying number of frogs denoted by $N(t)$. The  loss rate, $\mu$, is independent of environment.

Based on \deidentify{our earlier laboratory experiment}{the laboratory data from the study by Waddle \etal} \cite{waddle24}, we divided a frog's infectious period into three discrete infectious states ($I_{k,j}$ where $k\in\{U,V\}$, $j\in{\{1,2,3}\}$) to capture the time-varying infectiousness of a chytridiomycosis infection. 
The importance of including information on changing infectiousness as disease progresses has been identified in previous models \cite{hollanders2023recovered}. 
The times to transition between these three infection states ($1/\gamma_1$ and $1/\gamma_2$) and recovery back to susceptible ($1/\omega$) are the same for unvaccinated and vaccinated frogs. 

The force of infection, $\lambda_h(t)$, for a given environment $h$ is given by, 

\begin{align}
  \lambda_h(t) &= \beta_h \left(m_{U,1}I_{U,1} + m_{V,1}I_{V,1}+m_{U,2}I_{U,2} + m_{V,2}I_{V,2} + m_{U,3}I_{U,3} +m_{V,3}I_{V,3}   \right),  \nonumber \\
  &= \beta_h \sum_{k\in\{U,V\}} \sum_{j=1}^3 m_{k,j}I_{k,j},
  \label{eqn:foi}
\end{align}

where $\beta_h$ is the transmissibility of chytridiomycosis infection in environment $h$, $h \in$ \{shaded (\textit{sh}), unshaded (\textit{un})\}. 
It is known that the intensity of a frog's infection is correlated with their rate of shedding, and so the force of infection includes relative infectiousness terms $m_{k,j}$, which depend on the frog's infectious time-course and vaccination status prior to the current infection.
We fixed $m_{U,3}=1$, meaning that each other $m_{k,j}$ can be interpreted as the \textit{relative} infection intensity of compartment $I_{k,j}$ relative to $I_{U,3}$. 

The susceptibility of vaccinated frogs relative to unvaccinated frogs is incorporated through the multiplier, $\alpha$, on the transmissibility parameter, as shown in \cref{fig:model-diagram}.
Models for the shaded and unshaded environments are independent.
We assume only the transmissibility ($\beta_h$ in \cref{eqn:foi}) changes between environments to account for the temperature dependence of the infection load.
Within each mesocosm (independent experiment), it is assumed that frogs mix homogeneously, and there is no interaction in frogs between mesocosms.

\subsubsection{Initial conditions} \label{sec:initial-conditions}
In the Waddle \textit{et al.} \cite{waddle24} study, each mesocosm (independent experiment) started with 20~frogs.
Half of these frogs had been previously infected and then had their infection cleared prior to the experiment (vaccinated). 
The other half had no previous infection (unvaccinated).
Of the 10~frogs from each vaccination arm, half were then infected at the beginning of the experiment. This was the same between the two environmental conditions. Therefore, in each mesocosm, the initial conditions were: $S_U = 5$, $I_{U,1} = 5$, $S_{V} = 5$, $I_{V,1} = 5$. 
All other model compartments had an initial condition of 0.

\begin{table}
    \centering
    \begin{tabularx}{0.8\linewidth}{
    >{ \hsize = .65\hsize } C 
    >{ \hsize = .35\hsize } C 
}
\toprule
    Transition & Rate \\ 
\midrule
    $(S_U, I_{U,1}) \rightarrow (S_U-1, I_{U,1} + 1)$ & $\lambda_h(t) S_U (N(t)-1)^{-1}$ \\
    $(S_V, I_{V,1}) \rightarrow (S_V-1, I_{V,1} + 1)$ & $\alpha\lambda_h(t) S_V (N(t)-1)^{-1}$ \\
    $(I_{k,1}, I_{k,2}) \rightarrow (I_{k,1}-1, I_{k,2}+1) \quad k \in \{U,V\}$ & $\gamma_1 I_{k,1}$ \\
    $(I_{k,2}, I_{k,3}) \rightarrow (I_{k,2}-1, I_{k,3}+1) \quad k\in \{U,V\}$ & $\gamma_2 I_{k,2}$ \\
    $(I_{k,3}, S_V) \rightarrow (I_{k,3}-1, S_V+1) \quad k\in \{U,V\}$ & $\omega I_{k,3}$ \\
    $X \rightarrow X-1 \quad \forall X $ & $\mu X$ \\
\bottomrule
\end{tabularx}
    \caption{Transitions for the CTMC model for chytridiomycosis infection dynamics. $N(t)$ is the sum of frog counts across all compartments at time $t$. Subscript $k$ denotes vaccination status, where $U$ corresponds to unvaccinated (no prior infection), and $V$ corresponds to vaccinated (prior infection). Subscript $h$ denotes environment (shaded or unshaded). The final transition (frog loss) occurs $\forall X \in \{S_k, I_{k,j}\; | \; k=\{U,V\}, \; j = \{1, 2, 3\} \}$.
    }
    \label{tab:transition-rates}
\end{table}

\subsection{Data}
The data used for this study is a combination of both laboratory and outdoor mesocosm data from Waddle \textit{et al.} \cite{waddle24}.
Both laboratory and outdoor experiments were performed using  the native Australian frog species, the green and golden bell frog (\textit{Litoria aurea}).
The laboratory data was used to inform infectiousness and the duration of infectious periods.
The outdoor mesocosm data was used to fit the CTMC model.
The fitting process is described in detail in \cref{sec:abc}.

\subsubsection{Laboratory data} \label{sec:lab_data}
Longitudinal laboratory data on infection intensity over time was used to divide a frog's infectious period into three distinct periods \cite{waddle24}. Using this data, frogs were considered to be the most infectious when their infection intensity was above $10^{5}$ and $10^{4}$ for unvaccinated and vaccinated frogs, respectively, corresponding to the $I_{k,2}$ state for $k \in \{U,V\}$. Infection intensity values below these thresholds may correspond to the initial rise in intensity following infection $I_{k,1}$, or the decline in intensity on the path to complete loss of infectiousness $I_{k,3}$, discussed in \cref{sec:methods-frogassignment}. This same data was also used to inform the infectiousness in each of the infectious compartments relative to $I_{U,3}$, based on the intensity of infections observed in unvaccinated and vaccinated frogs (see \cref{tab:parameter-definitions} in Supplementary Information).

The laboratory data in Waddle \textit{et al.} \cite{waddle24} was also used to inform the transition times between infection states $I_{k,1}$, $I_{k,2}$ and $I_{k,3}$ for $k \in \{U,V\}$. Due to the considerable overlap in the weekly time course of infection intensities between vaccinated and unvaccinated laboratory frogs reported in \cite{waddle24}, we assumed the transition times between successive infection states were the same for unvaccinated and vaccinated frogs.  Based on this work, we set the average time spent in $I_{k,1}$, $1/\gamma_1$, to 2.5 weeks, and the average time spent in $I_{k,2}$, $1/\gamma_2$, to 4.5 weeks.

\subsubsection{Outdoor mesocosm data} \label{sec:methods-frogassignment}

\deidentify{Eight}{In Waddle \etal, eight} independent experiments (mesocosms) were performed under shaded and unshaded environments (four experiments per environment) \deidentify{, as described in}{} \cite{waddle24}. 
The population in each mesocosm at the start of the experiment has been described in \cref{sec:initial-conditions}.
In order to fit the CTMC model using the outdoor mesocosm data, we needed to assign frogs to states at each time point. 
The initial state of each frog was known.
At each subsequent time point, we used the frog's infection intensity, $L_t$, at the current and previous time point to determine their updated state. 
The criteria used for this assignment are outlined in \cref{tab:criteria-infection-assignment}. 
The infection intensity threshold for the $I_{U,2}$ and $I_{V,2}$ states, $L_{I2}$ in \cref{tab:criteria-infection-assignment}, were $10^{5}$ and $10^{4}$, respectively, as per \cref{sec:lab_data}. 

An unvaccinated frog was flagged as having a previous infection when their infection intensity returned to zero after their first non-zero infection intensity, or if they were observed with an increasing or high load, having previously been observed with a decreasing load. 
Once a frog was flagged as having a previous infection, it transitioned to the vaccinated pathway, according to the model shown in \cref{fig:model-diagram}. 
For example, if an unvaccinated frog with a decreasing load at observation time $t_{i-1}$ (i.e. in $I_{U,3}$) was observed at time $t_i$ to have an increasing load or a high load, its assigned state at time $t_i$ would be $I_{V,1}$ or $I_{V,2}$, respectively.
We assume that the time between observations (one or more weeks) is long enough that any fluctuations in infection load over a single infection would not be observed.

\begin{table}[htp]
    \centering
    \begin{tabularx}{\linewidth}{
    >{ \hsize = .31\hsize } C 
    >{ \hsize = .19\hsize } X 
    >{ \hsize = .25\hsize } C 
    >{ \hsize = .25\hsize } C 
}

\toprule

Criterion & Description & Unvaccinated with no previous infection & Vaccinated or previous infection \\

\midrule

$L_t=0$         & No infection &  $S_U$     &   $S_V$       \\
$L_t>L_{I2}$    & High load &   $I_{U,2}$   &   $I_{V,2}$    \\
$L_t \leq L_{I2}$ and $L_t \geq L_{t-1}$   &  Increasing load &  $I_{U,1}$   &   $I_{V,1}$    \\
$L_t \leq L_{I2}$ and $L_t<L_{t-1}$   & Decreasing load &   $I_{U,3}$   &   $I_{V,3}$    \\

\bottomrule

\end{tabularx}
    \caption{Criteria for assigning frogs to different infection compartments. Assignment of frogs to infection compartments is based on infection intensity $L_t$ at the current time point $t$ and at the previous time point ($t-1$). Infection state at $t=0$ is known. The infection intensity threshold $L_{I2}$ is set to $10^5$ and $10^4$ for unvaccinated and vaccinated frogs respectively.}
    \label{tab:criteria-infection-assignment}
\end{table}

\begin{figure}[h!]
    \centering
    \includegraphics[width=\textwidth]{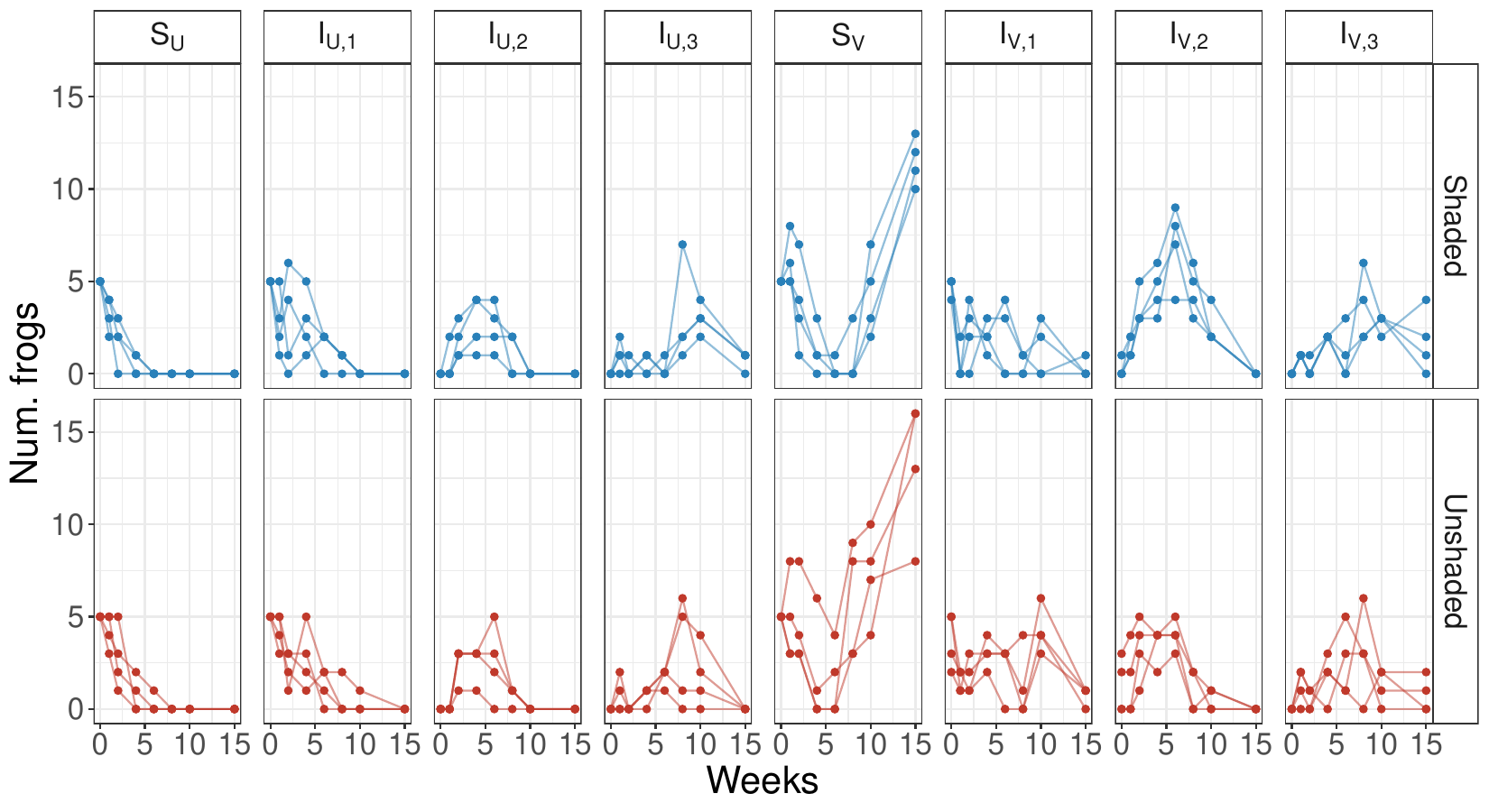}
    \caption{Experimental data used for the parameter estimation. Each line is a different mesocosm (independent experiment).}
    \label{fig:experimental-trajectories}
\end{figure}

Once each experimental frog was assigned its infection state at each time point, the data was aggregated at each time point for individual mesocosms to determine the number of frogs in each state for each mesocosm. 
In the experimental data, a frog was not always found at every time point.
A small number of these missing frogs were observed again at later time points, but the proportion of the missing observations was very small compared to the total number of observations (3.4\%) and hence we did not explicitly account for this in the model fitting process. The resulting state trajectories from each mesocosm are shown in \cref{fig:experimental-trajectories}.

\subsection{Parameter Estimation} \label{sec:abc}

We used an ABC rejection algorithm to estimate the free parameters of the model: $\theta = (\beta_{sh}$, $\beta_{un}$, $\alpha$, $\omega$). The ABC rejection algorithm draws a sample ($\theta$), from a defined set of prior distributions, simulates a model, which can be considered a `synthetic' experiment, and then accepts samples which have similar summary statistics to those calculated on the observed data \cite{Pritchard1999}. 

We simulated the CTMC model using the stochastic simulation algorithm, often referred to as the Gillespie algorithm \cite{Gillespie1977}. The model was implemented in R version~4.0.3 \cite{R}, using the \texttt{gillespiessa} package \cite{gillespiessa}.

In line with the experimental data, the model was simulated from $t=0$ to $t=15$ weeks, and `observed' at seven time points: $t_{\text{obs}} = \{1,2,4,6,8,10,15\}$.
The summary statistic, denoted $\Delta$, used was

\begin{equation}
\label{eqn:summary-stat}
  \Delta(\theta) = \sum_{x \in \Omega} \sum_{m=1}^8 \sum_{i=1}^7\left(x^\mathrm{sim}(t_i|\theta)-x^\mathrm{obs}_m (t_i)\right)^2,
\end{equation}

where $\Omega = \{S_k, \; I_{k,j} \; | \; k=\{U,V\}, \; j = \{1, 2, 3\} \}$. \Cref{eqn:summary-stat} can be interpreted as the sum of squared difference in the number of modelled and observed frogs at each time point ($t_i$), across each compartment ($x$), for each of the eight mesocosms ($m$). The prior distributions are given in \cref{tab:experimental-statistics}, and were chosen based on biological plausibility of the relevant parameter.
We applied a $0.2\%$ acceptance rate on $10^6$~samples from the prior distributions, resulting in 2000 accepted samples in the approximate posterior distribution.

\subsection{Code availability}
The data for the parameter estimation was processed from data available from Waddle \etal \cite{waddle24}.
The processed data and R code to reproduce the parameter estimation results are available from Zenodo \url{http://doi.org/10.5281/zenodo.15179344}.

\section{Results} \label{sec:results}

\subsection{Parameter identifiability and estimation}
We first tested the identifiability of parameters in our model in a simulation-estimation setting. 
Full details of the simulation-estimation experiment are in Supplement \ref{synthetic}.
We found we were unable to separately estimate the transmission coefficient, $\beta$, and the reduction in susceptibility due to vaccination, $\alpha$. However, the experiment showed that the product, $\alpha\beta$, was readily identifiable. 

The simulation-estimation study also showed that the loss rate of frogs, $\mu$, and the recovery rate, $\omega$, could not be reliably estimated from the available data simultaneously. As such, we calculate the loss rate using the average remaining count of frogs at the end of the experiment, which we denote as $\overline{N(T_f)}$ at the final observation time $T_f$.

The loss rate is estimated as,
\begin{equation} \label{eq:lossrate}
    \mu = -\frac{\mathrm{log}(\overline{N(T_f)}/N_0)}{T_f},
\end{equation}
for initial frog count $N_0=20$ and final time $T_f=15$~weeks. 
The average number of frogs found across all mesocosms at 15~weeks was 14.5. 
This gives an estimate of $\mu = 0.021$~per~week.

\begin{figure}[h!]
    \centering
    \begin{subfigure}{\linewidth}
        \caption{}
        \includegraphics[width=\linewidth]{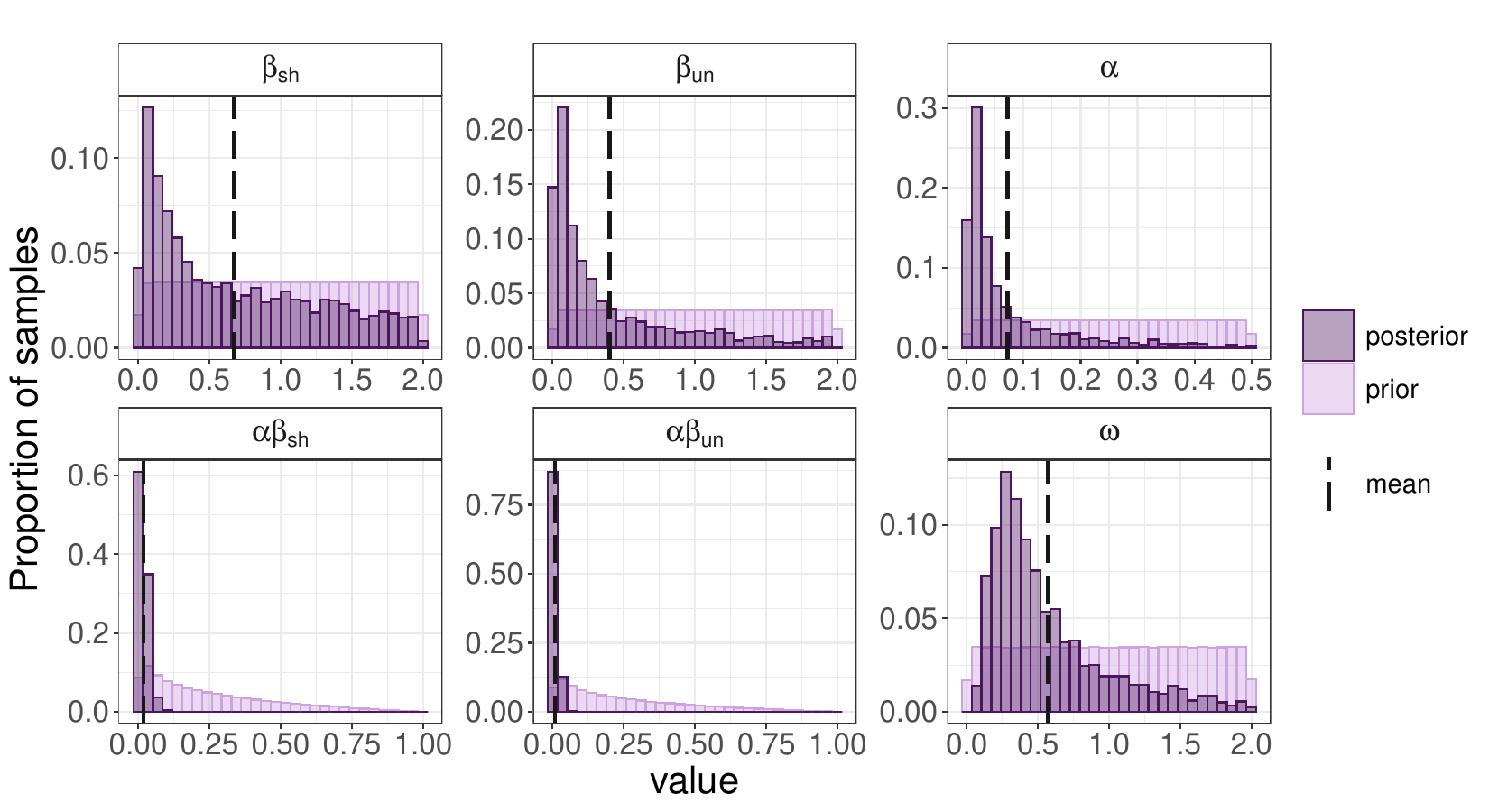}
        \label{fig:experimental-distributions}
    \end{subfigure}
    \begin{subfigure}{\linewidth}
        \caption{}
        \includegraphics[width=\linewidth]{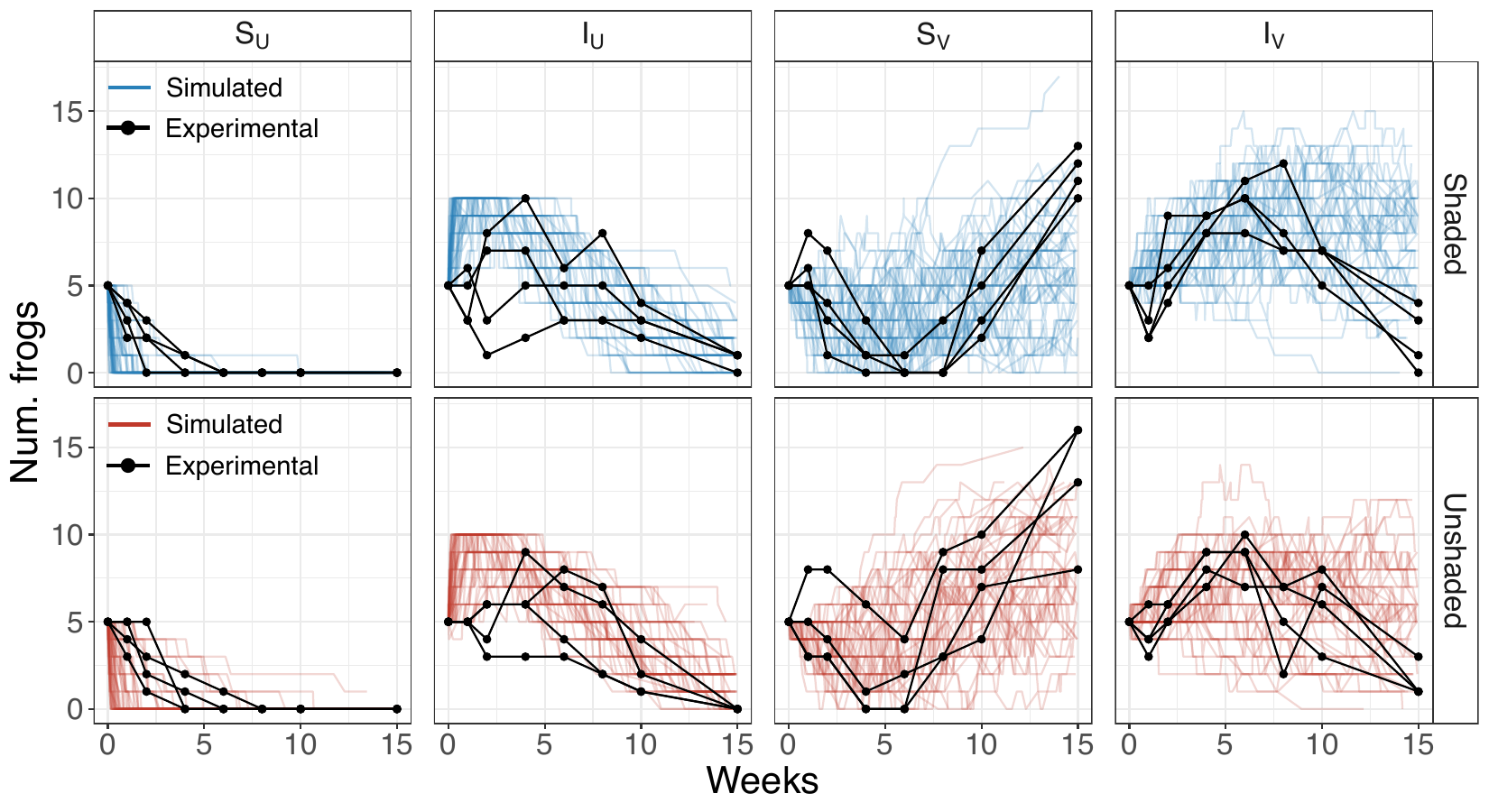}
        \label{fig:predictive-plots}
    \end{subfigure}
    \caption{
    (\subref{fig:experimental-distributions}) Prior (light purple) and posterior (dark purple) distributions with the mean of the posterior shown as a dashed line. We simulated $10^6$ samples from the prior and 0.2\% (2,000) were accepted.
    (\subref{fig:predictive-plots}) Example trajectories from 50~randomly sampled posterior parameter sets (blue/red lines), and experimental data (black lines). Here we plot the combined count of infected frogs for the unvaccinated and vaccinated cohort, i.e. $I_U = \sum_{j} I_{U,j}$ and $I_V= \sum_{j} I_{U,j}$ for $j=1,...,3$.
    }
\end{figure}

We obtained approximate posterior distributions for the four remaining parameters, $\theta=(\beta_{sh},\allowbreak \beta_{un},\allowbreak \alpha,\allowbreak \omega)$. 
The prior and posterior distributions are shown in \cref{fig:experimental-distributions}, and model fit is shown visually in \cref{fig:predictive-plots}.
The summary statistics of the distributions are given in \cref{tab:experimental-statistics}. 
Distributions shown on a log scale are also given in Supplement~\ref{si:log-distributions}.
Although the simulated trajectories from the accepted posterior samples show wide variance, the trends of the experimental data appear to be broadly captured by the model. 
The simulated trajectories are generally higher than  experimental data at $t=15$ weeks in both the $I_U$ and $I_V$ compartments. 
This indicates a potential underestimation of the recovery rate, $\omega$, or underestimation of the mortality rate, $\mu$.

\begin{table}[h!] \centering
    \begin{tabular}[t]{lcccc}
\toprule
Parameter           & Prior             & Posterior Mean & Posterior Median & Posterior 95\% CrI \\
\midrule                                
$\beta_\text{sh}$        & $ U(0,2)$     & 0.675 &   0.505   &  (0.025, 1.876) \\
$\beta_\text{un}$        & $ U(0,2)$     & 0.401 &   0.192   &  (0.008, 1.767) \\
$\alpha$            & $ U(0,0.5)$   & 0.072 &   0.029   &  (0.003, 0.373) \\
$\alpha\beta_\text{sh}$  &                   & 0.018 &   0.014   &  (0.001, 0.061) \\
$\alpha\beta_\text{un}$  &                   & 0.009 &   0.007   &  (0.000, 0.031) \\
$\omega$            & $ U(0,2)$     & 0.569 &   0.433   &  (0.122, 1.688) \\
\bottomrule
\end{tabular}
    \caption{Summary statistics for the posterior distributions. $U$ denotes the uniform distribution. Parameter descriptions are provided in Supplementary \cref{tab:parameter-definitions}.}
    \label{tab:experimental-statistics}
\end{table}

\subsection{Effects of environment and prior infection}
Using the parameter estimates in \cref{tab:experimental-statistics} we can quantify the effects of both environment and prior infection.
The protection offered by vaccination (prior infection) against chytridiomycosis infection is determined using the ratio between the mean of the $\alpha \beta_h$ and $\beta_h$ distributions for each environment ($h \in \{sh,un\}$):
\begin{equation}
    1-\frac{\alpha \beta_h}{\beta_h}.
\end{equation}
Our results show the vaccination effect is extremely strong, with a reduction in susceptibility for secondary infections of 97.3\% and 97.8\% for shaded and unshaded environments respectively. 

The effect of sunlight heating of the artificial refugia on chytridiomycosis transmission can be estimated by the ratio between distribution means of the transmission parameter, $\beta$, between the two environments ($\beta_{sh}$ and $\beta_{un}$):
\begin{equation}
    1 - \frac{\beta_{un}}{\beta_{sh}} .
\end{equation}
From \cref{tab:experimental-statistics}, we estimate a reduction in susceptibility due to sunlight heating to be 40.6\%.

\section{Discussion} \label{sec:discussion}

Ecological modelling studies of frog reintroduction programs illuminate challenges to the stability of frog populations due to chytridiomycosis \cite{stockwell2011}. Using a CTMC mathematical model and experimental data from Waddle \etal \cite{waddle24}, we quantified the effect of artificial habitat on chytridiomycosis transmission in green and golden bell frogs, as a potential pathway to supporting populations of wild frogs impacted by \bd. Our findings support the use of sunlight-heated artificial refugia as a method to assist with disease management: such refugia can reduce chytridiomycosis infection in frogs by approximately 40\%. 
Further, we have shown that prior \bd infection followed by recovery is highly protective: a relative reduction in susceptibility to infection of approximately $97\%$ was observed in frogs who had recovered from a prior infection. 

A previous study by Drawert \etal \cite{drawert2017using}, using a stochastic mathematical model of zoospore dynamics, found that a one-time antifungal protocol had the best outcome on frog population survival after an outbreak, compared to the other strategies of culling or disinfection of the environment. 
Our results also support the efficacy of this approach, and further show that recovered frogs under this scenario would experience a potent vaccination effect, increasing their resistance to disease and consequently the overall conservation benefits.
This vaccination effect could also improve frog population survival under endemic chytridiomycosis conditions, however future work is required to extend these results to larger populations, different species and longer time scales. 
We have limited our scope to the timescale and population size of the experimental data --- such extensions would require coupling the demography of the frog population to the disease dynamics, potentially using a similar approach to Briggs \etal \cite{briggs05}. 
Extension of these results to species of frogs other than green and golden bell frogs will require further experimental data, due to between species differences in susceptibility to \bd at warmer temperatures and resistance after clearance of infection \cite{cashins13,cohen17}.

Our modelling approach incorporated a method to account for varying infection intensity in the force of infection, through multiple infection compartments, which is considered a key component of chytridiomycosis dynamics \cite{hollanders2023recovered}.
While previous studies have observed an association between infection intensity and mortality \cite{hollanders2023recovered,carey06}, we did not observe such an effect in Waddle \etal 's \cite{waddle24} experimental data, used in our model. The authors of Waddle \etal \cite{waddle24} designed the experiment to limit the time frogs were exposed to temperatures that favored chytridiomycosis, running the experiment from mid winter into spring where warm temperatures could temper infections before  wide-scale mortality occurred. Due to this experimental design we did not incorporate a dependence of mortality on infection intensity in our model.
We limit our scope here to the effect of sunlight-heating and vaccination on disease transmission, and leave the study of mortality dependence on sunlight-heating to future work when appropriate data becomes available.

A particular strength of our study is the use of a CTMC model, with estimation performed in a Bayesian framework using an approximate Bayesian computation (ABC) rejection algorithm. The Markov chain model construction allows the results to be robust to numerical issues when the number of frogs becomes small. The Bayesian framework allows for estimation of a full posterior distribution, giving robust estimates of uncertainty around our parameters of interest. 

One of the main challenges in this study was the sample size of the data. Each of the eight mesocosms began with 20 frogs and the system was observed at seven time points over a period of 15 weeks. This limited observation frequency means it is possible that some transitions were not observed or were mis-identified in the data. This may explain why, at late observation time points, our simulations consistently overpredict the number of infected and underpredict the number of susceptible frogs in comparison to the experimental data. However, without additional experimental data, it is not possible to determine the true reason for this difference.

The small sample size led to significant uncertainty in our parameter estimates and meant that some parameters, such as the death rate, could not be estimated in the Bayesian framework. We were also unable to independently estimate the `vaccination' effect ($\alpha$) and the transmission parameter ($\beta$) and rather estimated these values using ratios of distribution means. The precise value of the vaccination effect, $\alpha$, is important for understanding the value of recovery from infection in the frogs, and is an important direction for future work. Our simulation-estimation study with the same characteristics as the experimental data provides transparency on the strength and limitations of our dataset and conclusions. 

This is the first mathematical modelling study that has focussed on habitat adjustment for frogs, as a way to combat chytridiomycosis infections. We provide estimation for the impact of sunlight-heated shelters on frog resistance to infections along with an estimation of the effect of exposure and recovery on subsequent susceptibility. Wider adoption of these cheap and easy to install shelters may lead to a reduction in chytridiomycosis prevalence amongst amphibians. With further work, this model could be used by wildlife managers to model the field deployment of refugia, and determine the level of refugia required in the landscape to ensure amphibian population recovery and stability. This future work will primarily include extending the model to incoporate frog population demography and infection-dependent mortality, and testing of the model and estimation approach in other settings and for other species.

\bibliographystyle{apalike}
\bibliography{references}

\begin{thebibliography}{}

\bibitem[Berger et~al., 1998]{berger1998}
Berger, L., Speare, R., Daszak, P., Green, D.~E., Cunningham, A.~A., Goggin, C.~L., Slocombe, R., Ragan, M.~A., Hyatt, A.~D., McDonald, K.~R., Hines, H.~B., Lips, K.~R., Marantelli, G., and Parkes, H. (1998).
\newblock Chytridiomycosis causes amphibian mortality associated with population declines in the rain forests of {{Australia}} and {{Central America}}.
\newblock {\em Proceedings of the National Academy of Sciences}, 95(15):9031--9036.

\bibitem[Berger et~al., 2004]{berger2004}
Berger, L., Speare, R., Hines, H., Marantelli, G., Hyatt, A., McDonald, K., Skerratt, L., Olsen, V., Clarke, J., Gillespie, G., et~al. (2004).
\newblock Effect of season and temperature on mortality in amphibians due to chytridiomycosis.
\newblock {\em Australian Veterinary Journal}, 82(7):434--439.

\bibitem[Briggs et~al., 2010]{briggs2010enzootic}
Briggs, C.~J., Knapp, R.~A., and Vredenburg, V.~T. (2010).
\newblock Enzootic and epizootic dynamics of the chytrid fungal pathogen of amphibians.
\newblock {\em Proceedings of the National Academy of Sciences}, 107(21):9695--9700.

\bibitem[Briggs et~al., 2005]{briggs05}
Briggs, C.~J., Vredenburg, V.~T., Knapp, R.~A., and Rachowicz, L.~J. (2005).
\newblock Investigating the {{Population-Level Effects}} of {{Chytridiomycosis}}: {{An Emerging Infectious Disease}} of {{Amphibians}}.
\newblock {\em Ecology}, 86(12):3149--3159.

\bibitem[Carey et~al., 2006]{carey06}
Carey, C., Bruzgul, J.~E., Livo, L.~J., Walling, M.~L., Kuehl, K.~A., Dixon, B.~F., Pessier, A.~P., Alford, R.~A., and Rogers, K.~B. (2006).
\newblock Experimental {{Exposures}} of {{Boreal Toads}} ({{Bufo}} boreas) to a {{Pathogenic Chytrid Fungus}} ({{Batrachochytrium}} dendrobatidis).
\newblock {\em EcoHealth}, 3(1):5--21.

\bibitem[Cashins et~al., 2013]{cashins13}
Cashins, S.~D., Grogan, L.~F., McFadden, M., Hunter, D., Harlow, P.~S., Berger, L., and Skerratt, L.~F. (2013).
\newblock Prior infection does not improve survival against the amphibian disease {{Chytridiomycosis}}.
\newblock {\em PloS One}, 8(2):e56747.

\bibitem[Cohen et~al., 2017]{cohen17}
Cohen, J.~M., Venesky, M.~D., Sauer, E.~L., Civitello, D.~J., McMahon, T.~A., Roznik, E.~A., and Rohr, J.~R. (2017).
\newblock The thermal mismatch hypothesis explains host susceptibility to an emerging infectious disease.
\newblock {\em Ecology Letters}, 20(2):184--193.

\bibitem[Drawert et~al., 2017]{drawert2017using}
Drawert, B., Griesemer, M., Petzold, L.~R., and Briggs, C.~J. (2017).
\newblock Using stochastic epidemiological models to evaluate conservation strategies for endangered amphibians.
\newblock {\em Journal of the Royal Society Interface}, 14(133):20170480.

\bibitem[Gillespie, 1977]{Gillespie1977}
Gillespie, D.~T. (1977).
\newblock Exact stochastic simulation of coupled chemical reactions.
\newblock {\em The Journal of Physical Chemistry}, 81(25):2340--2361.

\bibitem[Hollanders et~al., 2023]{hollanders2023recovered}
Hollanders, M., Grogan, L.~F., Nock, C.~J., McCallum, H.~I., and Newell, D.~A. (2023).
\newblock Recovered frog populations coexist with endemic batrachochytrium dendrobatidis despite load-dependent mortality.
\newblock {\em Ecological Applications}, 33(1):e2724.

\bibitem[Pineda-Krch and Cannoodt, 2022]{gillespiessa}
Pineda-Krch, M. and Cannoodt, R. (2022).
\newblock {\em GillespieSSA: Gillespie's Stochastic Simulation Algorithm (SSA)}.
\newblock R package version 0.6.2.

\bibitem[Piotrowski et~al., 2004]{piotrowski2004}
Piotrowski, J.~S., Annis, S.~L., and Longcore, J.~E. (2004).
\newblock Physiology of {{Batrachochytrium}} dendrobatidis, a chytrid pathogen of amphibians.
\newblock {\em Mycologia}, 96(1):9--15.

\bibitem[Pritchard et~al., 1999]{Pritchard1999}
Pritchard, J.~K., Seielstad, M.~T., Perez-Lezaun, A., and Feldman, M.~W. (1999).
\newblock Population growth of human y chromosomes: a study of y chromosome microsatellites.
\newblock {\em Molecular Biology and Evolution}, 16(12):1791--1798.

\bibitem[Puschendorf et~al., 2011]{puschendorf11}
Puschendorf, R., Hoskin, C.~J., Cashins, S.~D., Mcdonald, K., Skerratt, L.~F., Vanderwal, J., and Alford, R.~A. (2011).
\newblock Environmental {{Refuge}} from {{Disease-Driven Amphibian Extinction}}.
\newblock {\em Conservation Biology}, 25(5):956--964.

\bibitem[{R Core Team}, 2022]{R}
{R Core Team} (2022).
\newblock {\em R: A Language and Environment for Statistical Computing}.
\newblock R Foundation for Statistical Computing, Vienna, Austria.
\newblock R version 4.0.3.

\bibitem[Scheele et~al., 2019]{scheele2019}
Scheele, B.~C., Pasmans, F., Skerratt, L.~F., Berger, L., Martel, A., Beukema, W., Acevedo, A.~A., Burrowes, P.~A., Carvalho, T., Catenazzi, A., la~Riva, I.~D., Fisher, M.~C., Flechas, S.~V., Foster, C.~N., {Fr{\'i}as-{\'A}lvarez}, P., Garner, T. W.~J., Gratwicke, B., Guayasamin, J.~M., Hirschfeld, M., Kolby, J.~E., Kosch, T.~A., Marca, E.~L., Lindenmayer, D.~B., Lips, K.~R., Longo, A.~V., Maneyro, R., McDonald, C.~A., MendelsonIII, J., {Palacios-Rodriguez}, P., {Parra-Olea}, G., {Richards-Zawacki}, C.~L., R{\"o}del, M.-O., Rovito, S.~M., {Soto-Azat}, C., Toledo, L.~F., Voyles, J., Weldon, C., Whitfield, S.~M., Wilkinson, M., Zamudio, K.~R., and Canessa, S. (2019).
\newblock Amphibian fungal panzootic causes catastrophic and ongoing loss of biodiversity.
\newblock {\em Science}.

\bibitem[Stockwell et~al., 2011]{stockwell2011}
Stockwell, M., Clulow, S., Clulow, J., and Mahony, M. (2011).
\newblock The impact of the {{Amphibian Chytrid Fungus Batrachochytrium}} dendrobatidis on a {{Green}} and {{Golden Bell Frog Litoria}} aurea reintroduction program at the {{Hunter Wetlands Centre Australia}} in the {{Hunter Region}} of {{NSW}}.
\newblock {\em Australian Zoologist}, 34(3):379--386.

\bibitem[Voyles et~al., 2017]{voyles2017}
Voyles, J., Johnson, L.~R., Rohr, J., Kelly, R., Barron, C., Miller, D., Minster, J., and Rosenblum, E.~B. (2017).
\newblock Diversity in growth patterns among strains of the lethal fungal pathogen {{Batrachochytrium}} dendrobatidis across extended thermal optima.
\newblock {\em Oecologia}, 184(2):363--373.

\bibitem[Waddle et~al., 2024]{waddle24}
Waddle, A.~W., Clulow, S., Aquilina, A., Sauer, E.~L., Kaiser, S.~W., Miller, C., Flegg, J.~A., Campbell, P.~T., Gallagher, H., Dimovski, I., Lambreghts, Y., Berger, L., Skerratt, L.~F., and Shine, R. (2024).
\newblock Hotspot shelters stimulate frog resistance to chytridiomycosis.
\newblock {\em Nature}, 631(8020):344--349.

\bibitem[Waddle et~al., 2021]{waddle2021}
Waddle, A.~W., Rivera, R., Rice, H., Keenan, E.~C., Rezaei, G., Levy, J.~E., Vasquez, Y.~S., Sai, M., Hill, J., Zmuda, A., et~al. (2021).
\newblock Amphibian resistance to chytridiomycosis increases following low-virulence chytrid fungal infection or drug-mediated clearance.
\newblock {\em Journal of Applied Ecology}, 58(10):2053--2064.

\end{thebibliography}

\newpage
\begin{titlepage}
    \centering
    \vspace*{\fill}
    {\Huge \bfseries Supplementary Information \\[1em]}
    \vspace*{\fill}
\end{titlepage}

\appendix
\renewcommand{\thesubsection}{S\arabic{subsection}} 
\renewcommand{\thefigure}{S\arabic{figure}}   
\renewcommand{\thetable}{S\arabic{table}}     
\renewcommand{\theequation}{S\arabic{equation}}     
\setcounter{figure}{0}
\setcounter{table}{0}
\setcounter{equation}{0}

\subsection{Parameter table}
\begin{table}[h!]
    \centering
    \begin{tabularx}{\linewidth}{
    >{ \hsize = .14\hsize } C 
    >{ \hsize = .66\hsize } X 
    >{ \hsize = .20\hsize } C 
}

\toprule

Parameter  &  Meaning  &  Value \\

\midrule

$\beta_{sh}$ & 
    transmission coefficient for frogs in shaded mesocosms & 
    fitted \\
    
$\beta_{un}$ & 
    transmission coefficient for frogs in unshaded mesocosms & 
    fitted \\
    
$\alpha$  & 
    relative susceptibility to infection in vaccinated c.f. unvaccinated frogs & 
    fitted \\

$\omega$ & 
    transition rate   $I_{k,3}$ to $S_V$, $k=U,V$  & 
    fitted \\

$\mu$ & 
    loss rate of frogs & 
    0.021 \\

$m_{U,1}$ & 
    infectiousness in  $I_{U,1}$ relative to  $I_{U,3}$ & 
    10* \\

$m_{U,2}$ & 
    infectiousness in  $I_{U,2}$ relative to  $I_{U,3}$ & 
    100* \\

$m_{V,1}$ & 
    infectiousness in  $I_{V,1}$ relative to  $I_{U,3}$ & 
    1* \\
    
$m_{V,2}$ & 
    infectiousness in  $I_{V,2}$ relative to  $I_{U,3}$ & 
    10* \\

$m_{V,3}$ & 
    infectiousness in  $I_{V,3}$ relative to  $I_{U,3}$ & 
    0.1* \\

$\gamma_1$ & 
    transition rate  $I_{k,1}$ to  $I_{k,2}$, $k=U,V$ & 
    1/2.5 per week* \\

$\gamma_2$ & 
    transition rate  $I_{k,2}$ to $I_{k,3}$,  $k=U,V$ & 
    1/4.5 per week* \\

\bottomrule

\end{tabularx}
    \caption{Definitions of model parameters. All rates are per week. * indicates estimation from laboratory data. Note that $m_{U,3}=1$ by definition.}
    \label{tab:parameter-definitions}
\end{table}

\newpage
\subsection{Synthetic data study}\label{synthetic}
\allowdisplaybreaks

\subsubsection{Synthetic data generation}
\begin{figure}[h!]
    \centering
    \includegraphics[width=\linewidth]{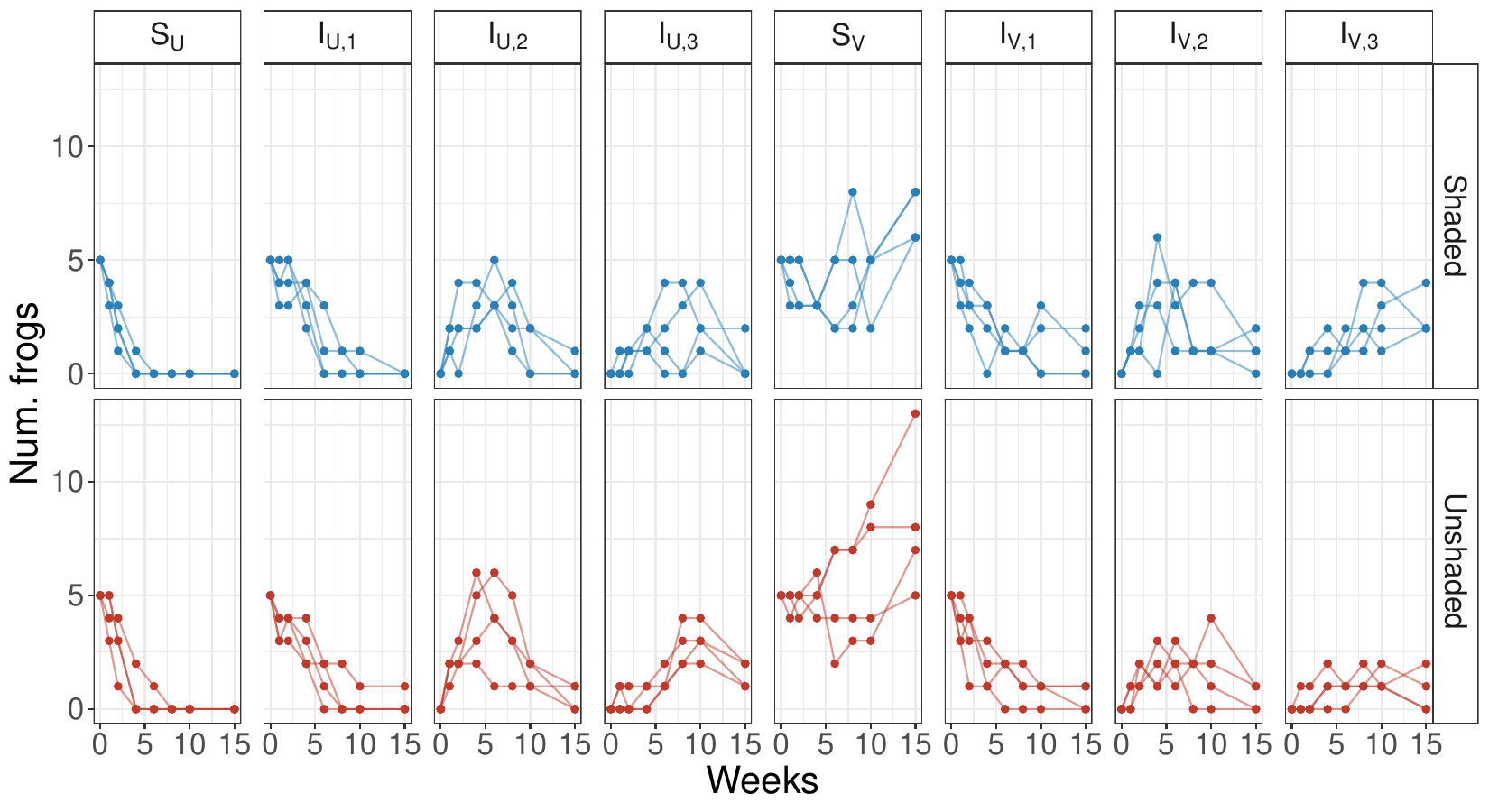}
    \caption{The synthetic trajectories generated to test the estimation method. We simulated four shaded and four unshaded trajectories using parameters estimated from fitting the data to a deterministic ODE model. }
    \label{fig:synthetic-trajectories}
\end{figure}

We used the GillespieSSA library \cite{gillespiessa} in R version~4.0.3 \cite{R} to simulate trajectories using parameters fit from the equivalent deterministic model (Section \ref{ode}), given in \cref{tab:deterministic-fits}.
We simulated four shaded and four unshaded environment trajectories to replicate the experimental data setup. 
The simulated trajectories are given in \cref{fig:synthetic-trajectories}. 

\begin{table}[h!]
\centering
    \begin{tabularx}{0.5\linewidth}{
    >{ \hsize = .3\hsize } C 
    >{ \hsize = .2\hsize } C 
    >{ \hsize = .5\hsize } C 
}

\toprule

Parameter & Value & 95\% CI \\

\midrule

$\beta_{sh}$ & 
    0.050 & (0.027, 0.072) \\
    
$\beta_{un}$ & 
    0.020 & (0.011, 0.029) \\
    
$\alpha$ & 
    0.177 & (0.087, 0.267) \\
    
$\omega$ & 
    0.258 & (0.173, 0.342) \\
    
$\mu$ & 
    0.029 & (0.025, 0.033) \\

\bottomrule

\end{tabularx}





    
    
    
    


    \caption{Parameters estimated using a deterministic ODE model. }
    \label{tab:deterministic-fits}
\end{table}

\subsubsection{Loss rate}
As with the experimental data, we calculate the loss rate using the average remaining count of frogs at the end of the experiment from \cref{eq:lossrate}.
From the synthetic trajectories, we have on average $\overline{N(T_f)}=12.1$~frogs remaining after at final time $T_f=15$~weeks. 
This gives an estimate of $\mu=0.033$~per week.

\subsubsection{Parameter estimation}

\begin{figure}[h!]
    \centering
    \begin{subfigure}{\linewidth} \caption{Linear scale}
    \includegraphics[width=\linewidth]{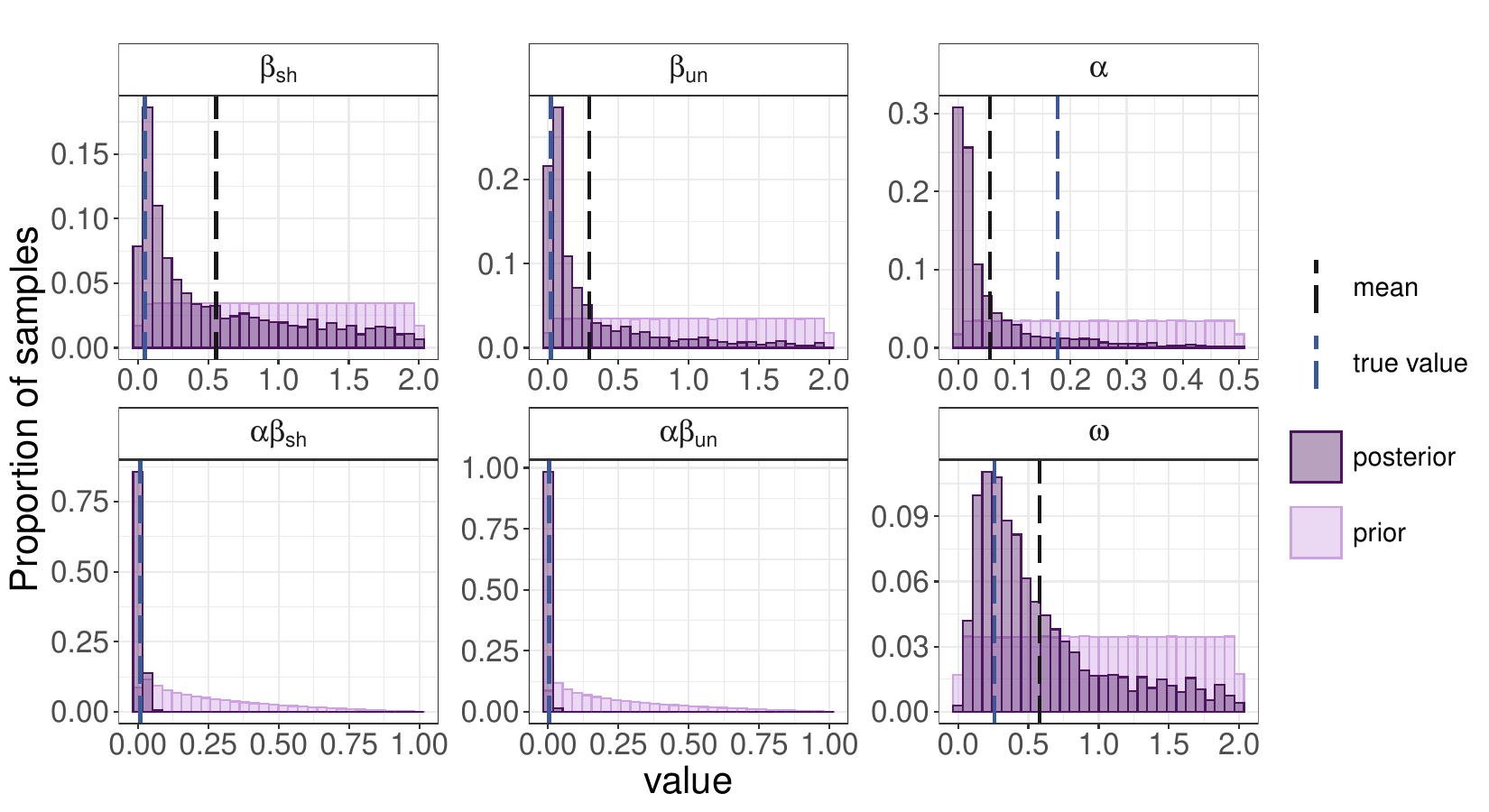}
    \end{subfigure}
    \begin{subfigure}{\linewidth} \caption{Log scale}
    \includegraphics[width=\linewidth]{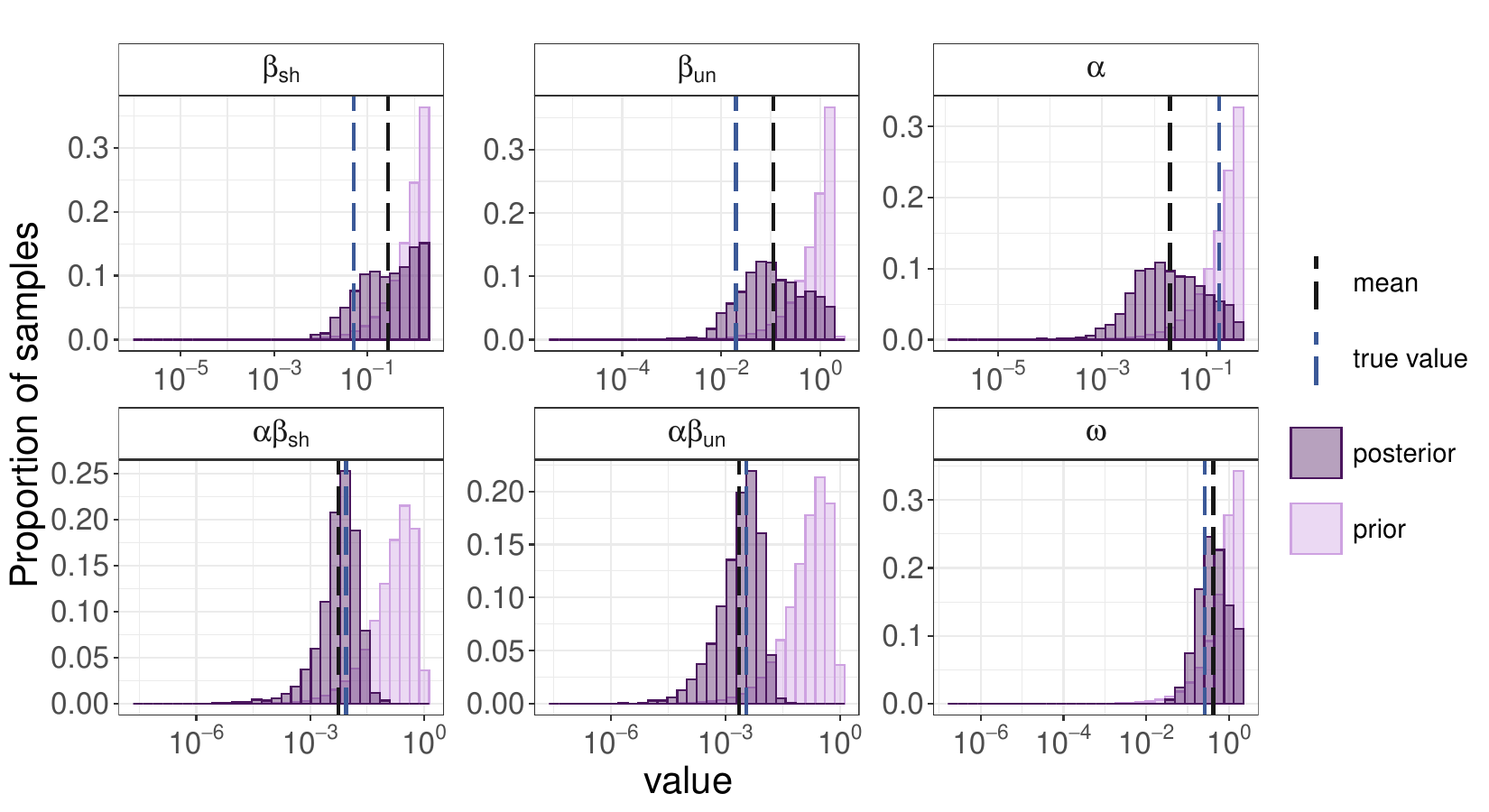}
    \end{subfigure}
    \caption{Comparison of prior and posterior distributions. True value is the value used to generate the synthetic data. Results shown on both linear scale (a) and log scale (b) for the parameter value.}
    \label{fig:synthetic-distributions}
\end{figure}

The prior and posterior distributions for the simulation study are shown in \cref{fig:synthetic-distributions}, with the summary statistics given in \cref{tab:synthetic-statistics}.

\begin{table}[h!] \centering
    \begin{tabular}[t]{lccccc}
\toprule
Parameter                & Prior             & True Value & Posterior Mean & Posterior Median & Posterior 95\%CrI \\
\midrule   
$\beta_\text{sh}$        & $ U(0,2)$     & 0.050 & 0.557     & 0.315     & (0.017, 1.845) \\
$\beta_\text{un}$        & $ U(0,2)$     & 0.020 & 0.294     & 0.103     & (0.008, 1.615) \\
$\alpha$                 & $ U(0,0.5)$   & 0.177 & 0.057     & 0.019     & (0.001, 0.330) \\
$\alpha\beta_\text{sh}$  &                   & 0.009 & 0.009     & 0.007     & (0.000, 0.032) \\
$\alpha\beta_\text{un}$  &                   & 0.004 & 0.004     & 0.003     & (0.000, 0.015) \\
$\omega$                 & $ U(0,2)$     & 0.258 & 0.581     & 0.422     & (0.080, 1.823) \\
\bottomrule
\end{tabular}
    \caption{Summary statistics for the posterior distributions for the synthetic data. True value is the value used to generate the synthetic data. We simulated $10^6$ samples and accepted 0.2\% (2,000). }
    \label{tab:synthetic-statistics}
\end{table}

\begin{figure}[h!]
    \centering\includegraphics[width=\linewidth]{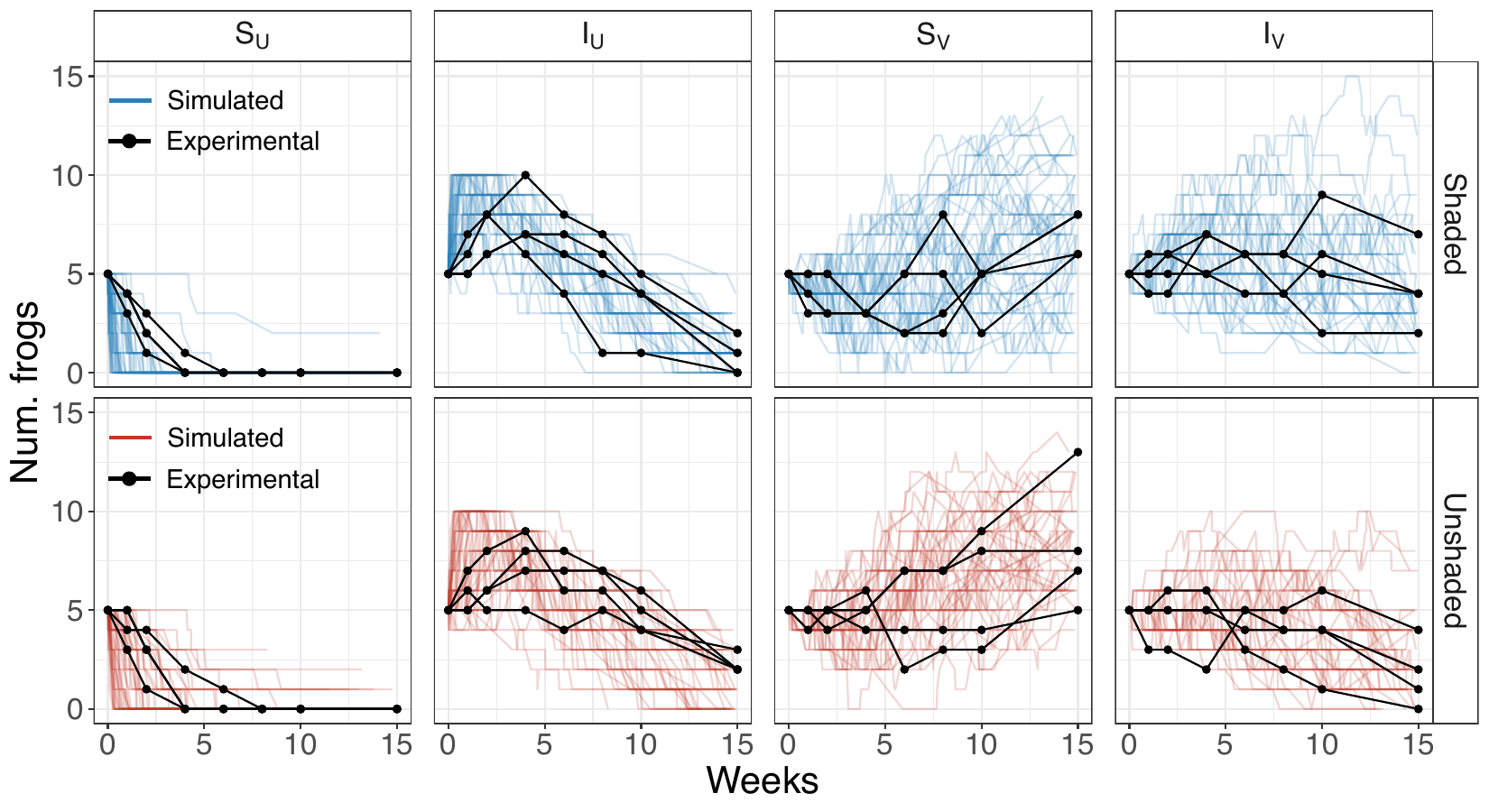}
    \caption{Example trajectories from 50~randomly sampled posterior parameter sets (blue/red lines) of the synthetic estimation, and synthetic data (black lines).}
    \label{fig:enter-label}
\end{figure}

\clearpage
\subsection{Ordinary differential equation model}\label{ode}

The number of frogs in each of the eight model compartments (Figure \ref{fig:model-diagram}) over time is governed by:

\begin{align}
\frac{dS_U}{dt}\  &= \  -  \frac{\lambda S_U }{N(t) - 1} - \mu S_U \\
\frac{dI_{U,1}}{dt}\  &=  \frac{\lambda S_U}{N(t) - 1}\  - \ \gamma_{1}I_{U,1} - \mu I_{U,1} \\
\frac{dI_{U,2}}{dt}\  &= \ \gamma_{1}I_{U,1}\  - \ \gamma_{2}I_{U,2} - \mu I_{U,2} \\
\frac{dI_{U,3}}{dt}\  &= \ \gamma_{2}I_{U,2\ } - \ \omega I_{U,3} - \mu I_{U,3} \\
\frac{dS_{V}}{dt}\  &= \  - \frac{\alpha \lambda S_{V}}{N(t) - 1} + \omega I_{U,3} + \omega I_{V,3} - \mu S_{V} \\
\frac{dI_{V,1}}{dt}\  &= \frac{\alpha \lambda S_{V}}{N(t) - 1}\  - \ \gamma_{1}I_{V,1} - \mu I_{V,1} \\
\frac{dI_{V,2}}{dt}\  &= \ \gamma_{1}I_{V,1\ } - \ \gamma_{2}I_{V,2} - \mu I_{V,2} \\
\frac{dI_{V,3}}{dt}\  &= \ \gamma_{2}I_{V,2}\  - \ \omega I_{V,3} - \mu I_{V,3}
\end{align}

where $N(t)$ is the total number of frogs in all compartments at
time point $t$, assumed to monotonically decrease at rate $\mu$ over
time. 
The force of infection is given by \cref{eqn:foi}.

\newpage
\subsection{Prior and posterior distributions for experimental data on log scale} \label{si:log-distributions}
\begin{figure}[h!]
    \centering
    \includegraphics[width=\linewidth]{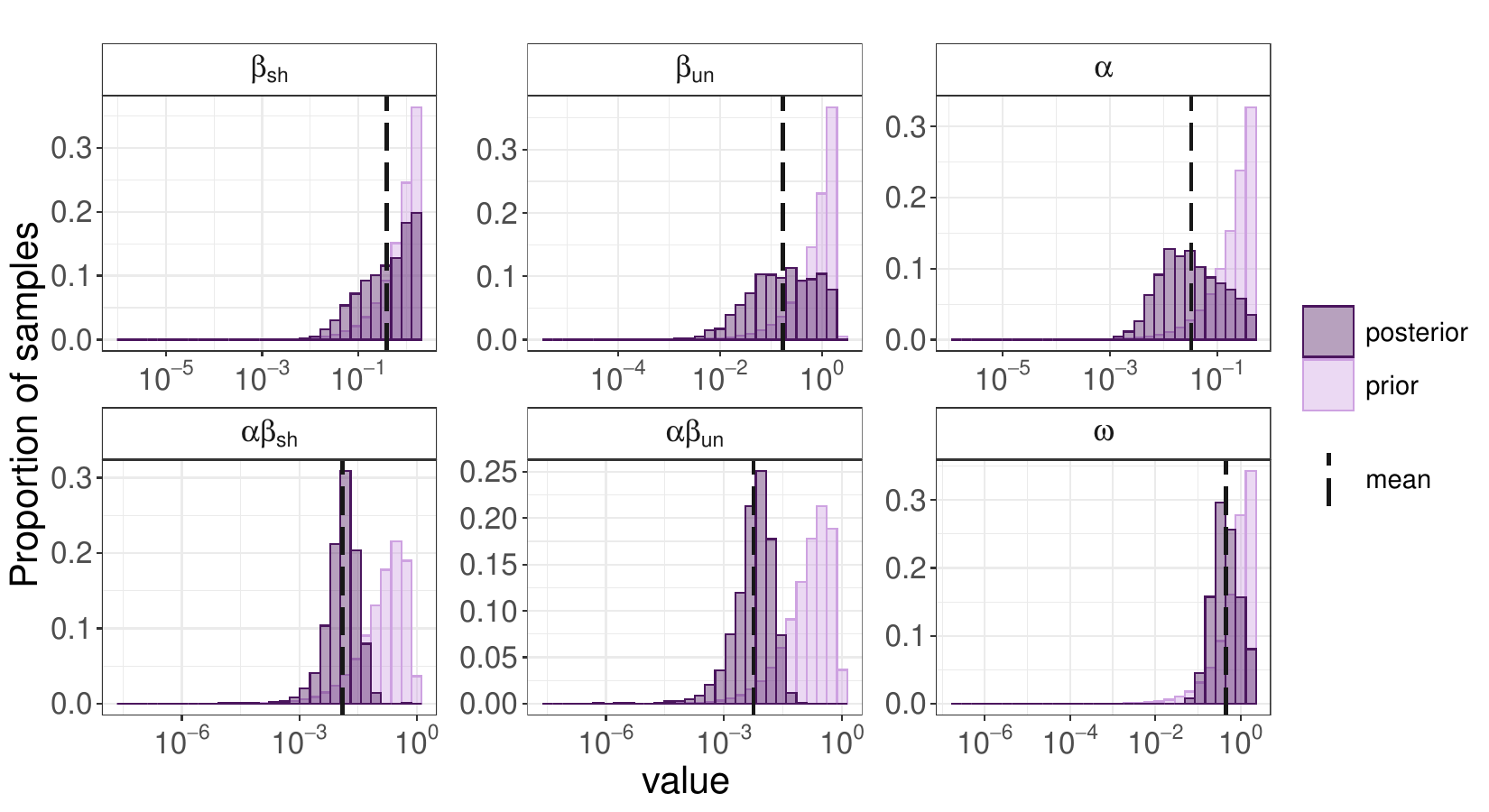}
    \caption{Prior and posterior distributions with the parameter on a log scale, to complete results shown in \cref{fig:experimental-distributions}.}
    \label{fig:experimental-distributions-log}
\end{figure}

\end{document}